\documentstyle[aps,multicol]{revtex}

\def \beq {\begin{equation}}
\def \eeq {\end{equation}}
\def \beqa {\begin{eqnarray}}
\def \eeqa {\end{eqnarray}}

\raggedcolumns

\begin{document}

\begin{center}
\vspace*{0.5cm}
{\large\bf  THE TWIST MODE IN ATOMIC FERMI GASES}\\[1.cm]

X. Vi\~nas$^1$, R. Roth$^2$, P. Schuck$^3$ and J. Wambach$^4$\\[0.3cm]

{\it $^1$Departament d'Estructura i Constituents de la Mat\`eria, }\\
{\it Facultat de F\'{\i}sica, Universitat de Barcelona,} \\
{\it Diagonal {\sl 647}, E-{\sl 08028} Barcelona, Spain} \\[6pt]

{\it $^2$Gesellschaft f\"ur Schwerionenforschung,} \\
{\it Planckstr. 1, D-{\sl 64291} Darmstadt, Germany} \\[6pt]

{\it $^3$Institut des Sciences Nucl\'eaires,} \\
{\it Universit\'e Joseph Fourier, CNRS--IN{\sl 2}P{\sl 3},} \\
{\it {\sl 53} Avenue des Martyrs, F-{\sl 38026} Grenoble-C\'edex, France}\\[6pt]

{\it $^4$Institut f\"ur Kernphysik,} \\
{\it Technische Universit\"at Darmstadt,} \\
{\it Schlossgartenstr.9, D-{\sl 64289} Darmstadt, Germany} \\[0.5cm]

PACS number(s): 03.75.Fi, 05.30.Jp

\end{center}

%
\vspace{0.5cm}

\begin{abstract}
The quantum-kinetic energy of a finite number of trapped fermionic
atoms provides a restoring force for shear motion due to a distortion
of the momentum distribution. In analogy to the twist mode of nuclear
physics it is proposed that counter-rotating the upper and lower
hemisphere of a spherical atomic cloud yields a finite-frequency mode
closely related to transverse zero-sound waves in bulk Fermi liquids.
\end{abstract}

\vspace{1cm}

%

\begin{multicols}{2}
The advent of Bose-Einstein condensation of trapped atomic $^{87}$Rb
in 1995 has initiated large experimental and theoretical activities in
the field of very dilute, almost ideal Bose gases \cite{R1,R2}. The
recent experimental achievements of trapping fermionic alkali-atoms
\cite{R3} raise hope that much progress will also be made in the near
future for Fermi gases. Indeed, Fermi-Dirac degeneracy of a mixture of
trapped $^{40}$K atoms in two different hyperfine states has been
achieved \cite{R4}. Recently, also the cooling of a mixture of bosonic
$^7$Li and fermionic $^6$Li atoms into a quantum degenerate regime was
accomplished \cite{Trus01}. On the theoretical side, intensive studies
have started as well. For instance, the possibility that certain gas
species show attractive interaction (e.g., $^{6}$Li) has initiated
studies on the possible superfluidity of such
systems~\cite{R5,R6}. The close analogy with another finite Fermi
system, the nucleus, has been pointed out~\cite{R6}. It is indeed
tempting to transpose many typical features of atomic nuclei to
trapped atomic Fermi gases. Besides the very spectacular superfluidity
properties there is interest in the spectrum of collective
excitations, most of them showing features proper to Landau's
zero-sound modes in bulk Fermi liquids~\cite{Baym}. For finite Fermi
systems zero sound translates into modes analogous to those of an
elastic body~\cite{Lamb,Bertsch}. One of the most remarkable examples
is the so-called 'twist mode'~\cite{Holz,R7,Schwes} in spherical
nuclei for which there is experimental evidence from backward
inelastic electron scattering~\cite{vNC}. This mode has also been
predicted to exist in medium to heavy spherical alkali metal clusters
as the most prominent magnetic multipole excitation~\cite{Nest}.  In a
macroscopic picture the twist mode comprises a coherent oscillation of
the particles in the upper half sphere versus those in the lower half
sphere.  For small amplitudes it corresponds to a purely kinetic
excitation without any spatial distortion of the equilibrium shape.
In this note we wish to investigate to what extent the twist mode may
also occur as a collective mode in very dilute, atomic Fermi gases in
the degenerate limit.

It is well known that the twist mode is generated by the operator~\cite{R7}
\beq
T = e ^{-i \alpha z l_z} = e^{ \alpha \vec{u}\cdot\vec{\nabla}} ,
\quad \vec{u}=(y z, -x z,0)
\label{eq1}
\eeq 
acting on the ground-state wave function of the Fermi system. As
is evident, $T$ induces a rotation of the particles around the
body-fixed $z$-axis with a rotation angle proportional to $z$, i.e.,
the rotation is clockwise for $z > 0$ and counterclockwise for $z <
0$. The amplitude of this twist is characterized by the angle $\alpha$.
One can verify that the twist corresponds to a magnetic mode of
spin-parity $J^\pi=2^{-}$. Although the operator (\ref{eq1}) induces
no change in the spatial distribution, the momenta become locally
distorted. The subsequent derivation of the mode frequency will
closely follow the original work of Holzwarth and
Eckart~\cite{R7}. For atomic Fermi gases with $N \simeq 10^{5}$ to
$10^{6}$ particles the Thomas-Fermi approach is very appropriate
\cite{R6} (as is the case of atomic bosons~\cite{R8}). Most relevant
for our purpose is the total kinetic energy of the system: 
\beq
E_{kin} (\alpha) = \int \frac{d^3rd^3p}{(2 \pi \hbar)^3} \frac{p^2}{2
m} f_{\alpha}(\vec{r},\vec{p})\, ,
\label{eq2} 
\eeq
where $f_{\alpha}$ is the distorted phase-space distribution in Thomas-Fermi
approximation
\beq
f_{\alpha} = \nu\, \theta \big( \tilde{p}_{F}^2(\vec{r},\hat{p})-p^2 \big)\, .
\label{eq3} 
\eeq
Here $\nu$ is the degeneracy factor and $\theta$ the unit step function.
The (quadrupole)deformed local momentum is given by
\beqa
\tilde{p}_F(\vec{r},\hat{p}) 
&=& p_F(\vec{r})\, N(\alpha)\, \big\{ 1 \nonumber \\ 
&-& \alpha \sqrt{\frac{2 \pi}{15}} 
\big[ y \big(Y_{21}(\hat{p})-
Y_{2-1}(\hat{p})\big) \\
& &\qquad\, + i\,x
\big( Y_{21}(\hat{p})+Y_{2-1}(\hat{p})\big)\big]\big\} \nonumber
\label{eq3a} 
\eeqa
where
\beq
p_F(\vec{r})=\sqrt{2m\big[\mu - V_{ex}(\vec{r}) 
  - g\, (\nu-1)\,\rho(\vec{r}) \big]}
\label{eq3b} 
\eeq
denotes the local Fermi momentum with chemical potential $\mu$ and
trapping potential $V_{ex}(\vec{r})$ and
\beq
N(\alpha)=1 - \alpha^2 \frac{x^2 +y^2}{15}\, .
\label{eq3b1} 
\eeq 
In the dilute gas limit it can be assumed 
that the two-body interaction is given by its long-wavelength limit
\beq
v(\vec{r} - \vec{r'}) = g\, \delta (\vec{r} - \vec{r'})
\label{eq3c} 
\eeq
with $g=4 \pi {\hbar}^2 a/m$, where $a$ is the s-wave scattering
length. For trapped atomic Fermions two situations can occur. For
example $^{40}K$ can be trapped as a mixture of atoms in two different
$m_F$ states, $m_F=9/2$ and $m_F=7/2$ \cite{R4}. In this case s-wave
scattering is realized and the interaction (\ref{eq3c}) is active. In
contrast, if the atoms are all in a single hyperfine state, there can
be no s-wave scattering. In the latter case p-wave interactions may
become very important \cite{Roth}. The inclusion of p-wave
interactions as well as the description of Fermion-Boson mixtures
requires a substantial extension of the fluid-dynamical formalism and
will be discussed in a following publication. In the present work we
concentrate on a two component Fermi gas with equal number of
particles in each magnetic substate. In this case the equilibrium
Thomas-Fermi equation for the density $\rho(\vec{r})$ of atoms in one
of the two $m_F$ states is given by \cite{Roth,R6}:
\beq 
\rho(\vec{r}) = \frac { p_F^3(\vec{r})}{6 \pi^2 \hbar^3}
\label{eq3d} 
\eeq
which leads to a cubic equation for $\rho$ which can be solved analytically.
To second order in $\alpha$ we then obtain
\beq
E_{kin}(\alpha) = \int\! d^3{r}\; \tau_{0}(\vec{r})\, \big[ 1 +
\frac{\alpha^2}{3}(x^2 +y^2) \big]
\label{eq4} 
\eeq
where $\tau_0$ is the total kinetic energy density at equilibrium
\beq
\tau_0(\vec{r}) = 2\,\frac{3}{5}\, \frac{\hbar^2}{2m}\,
\frac{p_F^5(\vec{r})}{6 \pi^2 \hbar^3}\, .
\label{eq5} 
\eeq
Since in lowest order, the potential energy contains no contribution from 
the twist mode, we obtain for the restoring force (assuming a spherically
symmetric trap of harmonic oscillator shape with frequency $\omega$) :
\beq
C = \frac{\partial^2 E_{kin}(\alpha)}{\partial \alpha^2} =
\frac{16 \pi}{9} \int\! dr\; r^4 \tau_0(r)
\label{eq6} 
\eeq
We also need to evaluate the mass parameter $B$ of the twist motion. As
usual in fluid dynamics it is given by
\beqa
B &=& m \int\! d^3{r}\; 2 \rho(\vec{r})\; u^2 \nonumber\\
&=& m \int\! d^3{r}\; 2 \rho(\vec{r})\; z^2(x^2+y^2) \nonumber\\
&=&\frac{16 \pi}{15} m \int\! dr\; r^6\, \rho(r)\, ,
\label{eq7} 
\eeqa
where $2 \rho(\vec{r})$ corresponds to the total density of
atoms. The twist frequency $\Omega_T$ is then obtained as
\beq
\hbar\Omega_{T} = \sqrt{\frac{C}{B}}\, .
\label{eq8} 
\eeq
We have considered two systems. One is $^{6}$Li with a very large
attractive scattering length of $a$=-2063$a_0$ ($a_0$ = Bohr radius)
\cite{R9}. The trapping potential was taken to be $\hbar \omega$ = 6.9
nK \cite{Trus01}. The other system is $^{40}$K with a repulsive
scattering length of $a$=157$a_0$ and $\hbar \omega$=1.6 nK
\cite{R10}. The results for the twist frequency as a function of the
particle number in each magnetic substate are given in Table 1. In
order to see how $\Omega_T$ depends on the interaction strength (which
may be variable due to the tuning of Feshbach resonances) we also list
$\Omega_T$ for various other values of the scattering lengths
(differing from the original ones by powers of 10).

From Table 1 it can be inferred that the influence of the interaction
on the twist frequency is very moderate. It is typically of the order
of 10$\%$ (except for very large particle number). This is consistent
with the expectation that, for transverse zero sound, s-wave
interactions give no contribution to the restoring
force~\cite{Schwes}.  The dependence of the twist frequency on the
interaction only enters through the mass parameter $B$ which depends
on the density (Eq.~\ref{eq7}) Depending on the sign of the
interaction the gas either expands (repulsive) or contracts
(attractive) relative to the free gas case.  Consequently the
frequency is decreased (increased) with respect to its non-interacting
value $\Omega_{T0} = \omega$ for repulsion (attraction).  This
feature should be measurable even though the absolute effect might be
small.

Very interesting possibilities arise when the p-wave interaction
becomes strong~\cite{Roth}. In this case there will be a significant
correction to the kinetic energy through the effective mass and hence
a large influence on the twist frequency. One may even encounter
instabilities, signaled by the exponential growth of the twist
amplitude.  Also, as mentioned above, one can then have a direct
influence of the interaction in a one component Fermi system.  Another
interesting issue is how the twist mode is influenced by eventual
superfluidity. It can be predicted that the twist mode ceases to exist
once pairing is strong enough for the system to reach its irrotational
flow limit~\cite{R6}. There may, however, be intermediate
situations. To our knowledge these possibilities have not been
addressed for the nuclear twist.
 
The question how to excite the twist mode in the experiment may not be
trivial. One could imagine utilizing the well-developed technique of
rotating trapped atoms~\cite{R11}. First a very elongated trap
potential is created. Subsequently a rotating laser field is wrapped
around the long axis inducing a rotation of the atomic cloud. If,
instead of applying the laser field parallel to the long axis, it is
incident at a certain angle $\phi$ and a mirror is placed parallel to
the long axis which reflects the laser beam in such a way that it hits
the other hemisphere at an angle $-\phi$, then the first hemisphere
will rotate in one direction and the other hemisphere in the opposite
direction (at least approximately) since the rotational sense of the
laser is inverted by the mirror. If the rotation is very gentle and
stopped at a certain time, the system will continue oscillating
(approximately) in the twist mode. Whether one can detect the twist
mode by switching off the trap potential and subsequently imaging the
velocity distribution of the expanding atoms remains an open question.

\section*{Acknowledgments}
\noindent
The above mentioned experimental possibility for exciting the twist
mode came up in a discussion with Y. Castin, J. Dalibard and
C. Salomon and is gratefully acknowledged. We also thank A. Richter
for fruitful discussions on the nuclear twist mode. One of us (X.V)
also acknowledges financial support from DGCYT (Spain) under grant
PB98-1247 and from the DGR (Catalonia) under grant 1998SGR-00011.

%

\end{multicols}
\clearpage
%
\vspace{2cm}

TABLE 1: The twist-mode frequencies $\Omega_T$ in units of the trap
frequency $\omega$ for several numbers of $^{6}$Li ($g<0$) and
$^{40}$K ($g>0$) atoms in each hyperfine state for different
scattering lengths (in units of the Bohr radius). The frequencies of
the harmonic oscillator traps are $\omega = 2\pi\times144$ Hz for 
$^6$Li and $\omega = 2\pi\times 33.5$ Hz for $^{40}$K, which 
corresponds to level spacings of $\hbar \omega$ = 6.9 nK and 1.6 nK, 
respectively.  

\vspace{0.2cm}
\begin{center}
\renewcommand{\arraystretch}{1.3}
\begin{tabular}{c | c c | c c | c c}
\hline\hline                  
$N$ & $\;\;a\;\;[a_0]\;\;$ & $\;\Omega_{T}\;\;[\omega]\;\;$ 
 & $\;\;a\;\;[a_0]\;\;$ & $\;\Omega_{T}\;\;[\omega]\;\;$ 
 & $\;\;a\;\;[a_0]\;\;$ & $\;\Omega_{T}\;\;[\omega]\;\;$ \\
\hline
$^{6}$Li &&&&&& \\
$\;1\times 10^3\;$  & -2063.0 & 1.042 & -206.3 & 1.004 & -20.63 & 1.000 \\
$5\times 10^3$  & -2063.0 & 1.056 & -206.3 & 1.005 & -20.63 & 1.000 \\
$1\times 10^4$  & -2063.0 & 1.064 & -206.3 & 1.006 & -20.63 & 1.001 \\
$5\times 10^4$  & -2063.0 & 1.087 & -206.3 & 1.007 & -20.63 & 1.001 \\
$1\times 10^5$  & -2063.0 & 1.101 & -206.3 & 1.008 & -20.53 & 1.001 \\
$2\times 10^5$  & -2063.0 & 1.116 & -206.3 & 1.009 & -20.63 & 1.001 \\
$5\times 10^5$  & -2063.0 & 1.142 & -206.3 & 1.011 & -20.63 & 1.001 \\
\hline\hline
$^{40}$K &&&&&& \\
$1\times 10^3$  & 157.0 & 0.996 & 15.7 & 1.000 & 1570.0 & 0.966 \\
$1\times 10^4$  & 157.0 & 0.995 & 15.7 & 0.999 & 1570.0 & 0.952 \\
$1\times 10^5$  & 157.0 & 0.992 & 15.7 & 0.999 & 1570.0 & 0.932 \\
$1\times 10^6$  & 157.0 & 0.989 & 15.7 & 0.999 & 1570.0 & 0.906 \\
$1\times 10^7$  & 157.0 & 0.984 & 15.7 & 0.998 & 1570.0 & 0.873 \\
\hline\hline
\end{tabular}
\end{center}


\end{document}